\documentclass{vgtc}                          

\graphicspath{{figures/}{pictures/}{images/}{./}} 

\usepackage{times}                     

\usepackage{tabu}                      
\usepackage{booktabs}                  
\usepackage{lipsum}                    
\usepackage{mwe}                       

\usepackage{mathptmx}                  

\onlineid{0}

\vgtccategory{Research}

\vgtcinsertpkg
\usepackage{hyperref} 

\usepackage[dvipsnames]{xcolor} 

\title{The Balance between Nuance and Clarity:\break 
Decluttering Tabular Sequential Graphs to Counter Money Laundering
}

\author{Salomé Esteves\thanks{e-mail: salome.esteves@feedzai.com}\\ %
        \scriptsize Feedzai %
\and Rita Costa\thanks{e-mail: rita.costa@feedzai.com}\\ %
     \scriptsize Feedzai %
\and Louise Fallon\thanks{e-mail: louise.fallon@mastercard.com}\\ %
     \scriptsize Mastercard %
\and Pedro Bizarro\thanks{e-mail: pedro.bizarro@feedzai.com}\\ %
     \parbox{1.4in}{\scriptsize \centering Feedzai}}

\teaser{
  \centering
  \includegraphics[width=\linewidth]{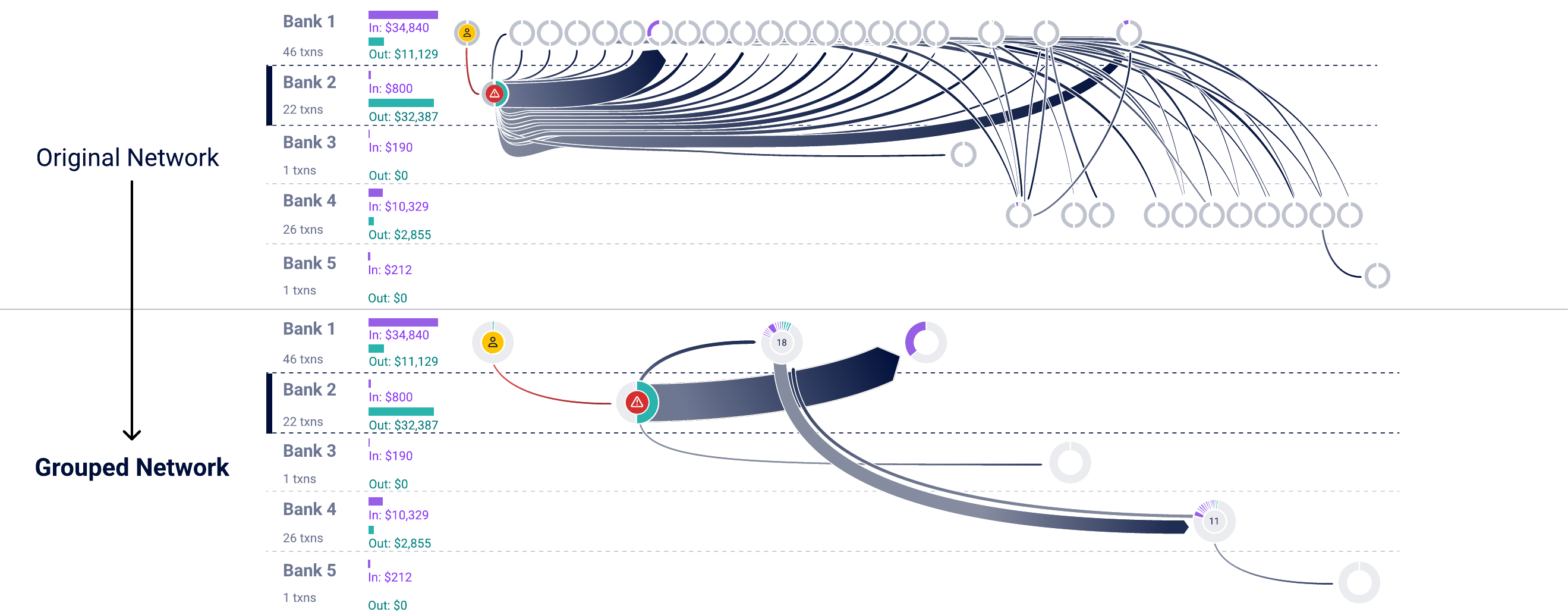}
  \caption{Example of a tabular sequential graph for a money laundering network before and after the implementation of a method of transaction amount-based node grouping (synthetic data used for privacy). Financial criminals try to spread and disguise their activities with 50 transactions, but node grouping reveals the flow clearly.}
  \label{fig:teaser}
}

\abstract{
    Money laundering is not only about moving illicit funds, but about hiding the money’s origin and traces to complicate detection. Financial criminals resort to many methods to avoid regulators and legal thresholds. But analysts investigating alerts, dedicated to pin mule accounts and track suspicious transactions daily, also have theirs. Network visualizations can be key in countering adversarial money laundering activities, especially if they provide a clear overview of the money flows and a seamless analysis experience, but they are often not structured for this type of task. That is why we propose a tabular sequential graph visualization tailored to money laundering analysis — following transactions (edges) from the victim account that triggered an alert through multiple accounts (nodes) and banks (rows). To reduce the number of nodes and edges, we propose three methods for grouping these tabular sequential graphs: an amount-based approach, a time-based approach, and a combined solution that considers both the transaction amount and its order. A qualitative user study with experts suggests that the most effective method in node reduction was not necessarily the most interesting for analysis and that there is a trade-off between manual work and time for interpretation in more granular graphs.

} 


\begin{document}


\firstsection{Introduction}

\maketitle

The number of nodes that a force-directed layout network can reach before it becomes unreadable is narrow. Research has shown that the readability of force-scattered networks can be compromised in graphs ranging from 20 to 100 nodes, depending on their density \cite{edge_trimming_2018, ghoniem_readability_2005, yoghourdjian_scalability_2021}. From this point on, interactions—such as panning, hovering, zooming, or motion—are useful to better the experience \cite{shneiderman_interactive_2013, ware_motion_2004}. But in our proposed tabular sequential graph (TSG), the threshold might be lower due to its constraints.

In a money laundering scheme, organized criminals perform a quick succession of transactions to avoid regulatory and legal thresholds and make the origin of the funds as untraceable as possible \cite{teichmann_recent_2020}. In this process of concealment, financial criminals might attempt multiple laundering methods, such as layering (passing money repeatedly to hide the origin of the funds), structuring (consecutive deposits and withdrawals), or smurfing (rapid transactions through multiple intermediaries) \cite{contreras_strategic_2026, halford_developing_2025, shadrooh_smotef_2024}.

Such schemes are highly volatile, with variable scope and duration across accounts and institutions. Their unpredictability has grown as online payments become faster, more convenient, and easier to anonymize. \cite{gilmour_reexamining_2022}.

In our case—a platform for cross-bank money-laundering tracking—it all starts with an alert. A victim has inadvertently transferred a sum of money to a potential mule account (a means to collect and move illicit funds) and the bank is notified. An analyst then starts an investigation to conduct two key analytical tasks:
Identify and pin accounts as mules - if these accounts are receiving illicit funds from victims or from other mule accounts.
Identify and pin transactions as scams, fraud, or mule-to-mule transactions.

In a scenario where financial criminals are actively trying to erase their traces, providing a visualization that prioritizes showcasing the links between accounts and the speed at which movements occur is key to counter money laundering activities.

Currently available solutions for analyzing fraud and money laundering \cite{noauthor_cambridge_nodate, noauthor_datawalk_nodate, noauthor_linkurious_nodate, noauthor_neo4j_2026} track the flow of money through networks and graph visualizations. However, these solutions often lack a view of how transactions occur over time and do not offer an alternative ordinal or sequential view. In short, most examples of networks applied to these use cases resort to force-directed layouts.

These network visualizations can expand and retract depending on the size of the dataset and literature has extensively covered decluttering methods for when force-directed layouts face an overwhelming accumulation of nodes and edges \cite{bach_towards_2017, balzer_level--detail_2007, edge_trimming_2018, shneiderman_interactive_2013, vehlow_visualizing_2017,  wang_visualizing_2019}.

However, traditional force-directed layouts did not meet our requirements since they could not ensure fixed space for each bank or convey a sense of time in the graph.
We propose a tabular sequential graph (TSG) in which the position of each node  (account) is fixed vertically on its row (bank). Nodes are placed on the x-axis according to a predefined order (time of first transaction) and never overlap across rows, showing the order in which the accounts entered the scheme and creating the columns in this tabular graph.

This leads to the main research question: How can we minimize the number of nodes and edges in tabular sequential graphs to their essential minimum to facilitate the two key analytical tasks that the platform supports - the identification of key accounts and transactions in money laundering analysis? The victim account sits on the far left by default, but flagged mule accounts appear throughout the graph.
Furthermore, we also inquire: What constitutes a large TSG in the context of money laundering detection in multi-bank networks? And in what circumstances must accounts and transactions stand alone in the graph and be excluded from aggregation?

We aim to reduce the number of nodes and edges in large TSG's to avoid additional intersections and improve clarity of high-value movements, fraudulent transactions, victim, and mule accounts.

Our main contributions are as follows: \begin{itemize}
    \item The tabular sequential graph (\hyperref[sec:tsg]{section 3}).
    \item Three node grouping algorithms for TSGs (\hyperref[sec:gs]{section 4}).
    \item  A targeted evaluation the grouping methods effects on two money laundering cases (\hyperref[sec:us]{section 5}).
    \item A meta-node that encodes accounts in a group (\hyperref[sec:tsg]{section 3}).
\end{itemize}

\section{Literature Review}
\label{sec:lr}

Graphs and networks -- used interchangeably in this paper -- are abstract, multivariate, and dynamic visualizations that display data points and the relations between them \cite{filipov_are_2023, saket_node_2014}. Although their complexity might vary, such displays can easily become cluttered due to a high-volume of overlapping nodes and intersecting edges. 

Severely populated graphs can negatively affect the user’s perception of the visualization, its readability and even its aesthetic considerations \cite{balzer_level--detail_2007, purchase_empirical_2002, shneiderman_interactive_2013}. 

Clutter-reduction strategies include edge-bundling \cite{bach_towards_2017} and edge-cutting  \cite{edge_trimming_2018} at edge-level; and filtering, clustering, node grouping and motif simplification \cite{shneiderman_interactive_2013} at node-level. However, these are tailored to force-directed networks, which lack the spatial and temporal restrictions of the layout we are proposing. 

Besides, while large networks can contain thousands or millions of data points \cite{filipov_are_2023}, a money laundering scheme becomes complex at a much smaller scale. In our environment, visualizations of a series of real-life scenarios showed that a case involving 15 accounts can become cluttered if all or most accounts belong to a single bank or to different banks. From that point forward, the goal is to reduce dispensable visual elements enabling the analyst to quickly process the information and proceed with the investigation. 

We adopt the node grouping definition by Schneidermann and Dunne \cite{shneiderman_interactive_2013}, by which nodes are aggregated into a single meta-node based on similar attributes, which leads to “a dramatic reduction in screen space required”. The meta-node then adopts a new visual representation to encode a group, through changes in color, glyphs or other visual attributes \cite{vehlow_visualizing_2017}. 

With added interaction, meta-nodes can also expand to reveal their initial state or any desired level of information \cite{wang_visualizing_2019}. This allows for the user to maintain a mental map of many levels of detail within the visualization and improve their understanding of the abstraction that occurs in clustering or aggregation methods \cite{balzer_level--detail_2007}.

Previous research has shown the usage of such aggregated nodes is useful in optimizing focus in network tasks in fraud detection \cite{feliciano_increasing_2022}. Aggregation methods on edge-level were not considered in this case because the original network already groups multiple transactions between two accounts into one single link. In other words, in our graph visualization any edge can be a meta-edge.

\section{The tabular sequential graph (TSG)}
\label{sec:tsg}

\begin{figure}
    \centering
    \includegraphics[width=1\linewidth]{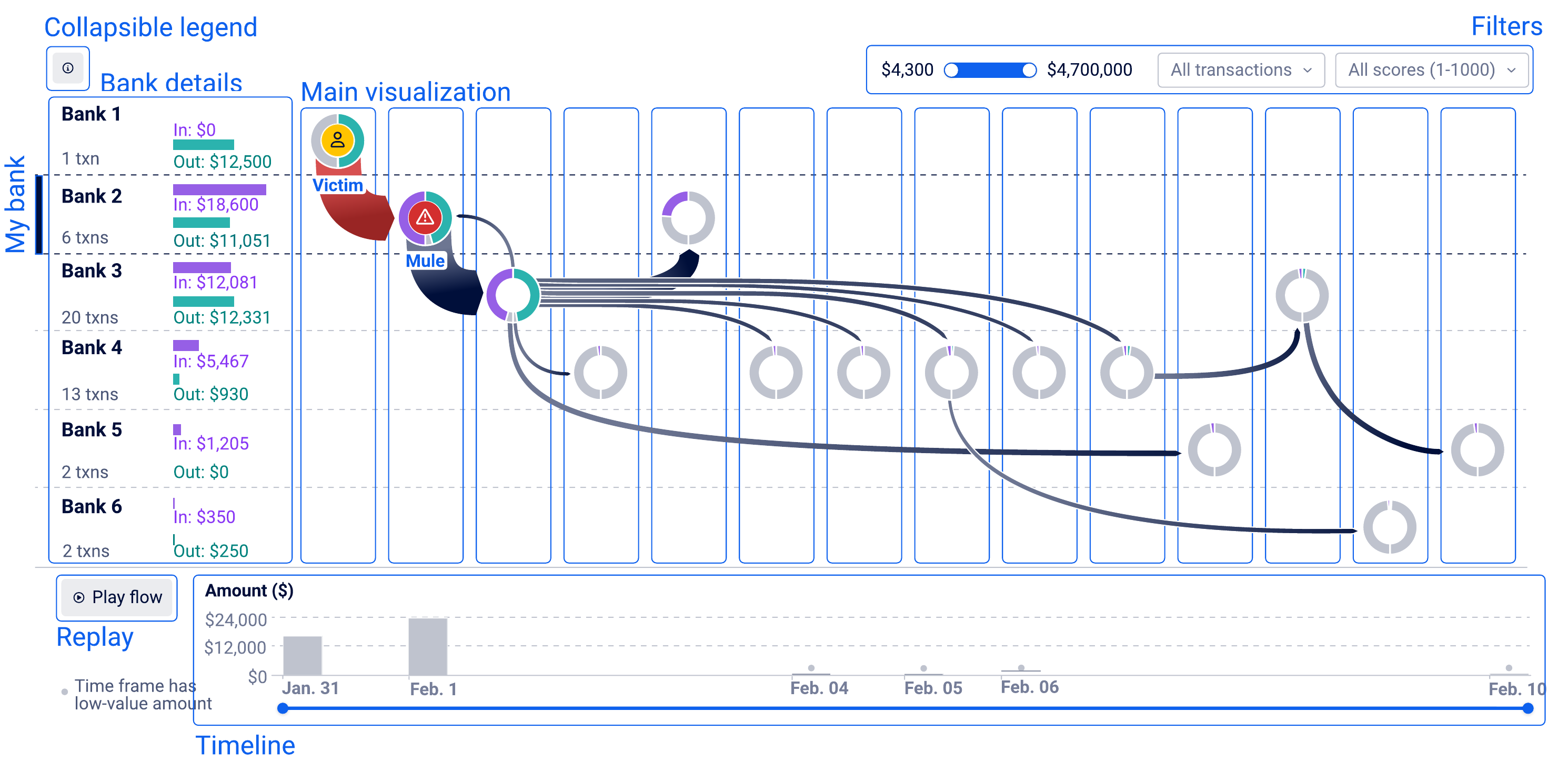}
    \caption{A tabular sequential network with six banks and fourteen accounts (synthetic data used for privacy).}
    \label{fig:network_15accounts}
\end{figure}

The TSG (Figure~\ref{fig:network_15accounts}) was tailored as a widget in a fraud analysis desktop-only platform, already deployed at the time of this research. It is displayed within a page where the analyst performs their investigation, along with additional alert details, such as information about the alerted account, KPIs regarding the network, and a history of all transactions made by the accounts in the scheme.

In this tabular structure, each bank in the network to has its own row within the graph, so the analyst can quickly perceive how many accounts (nodes) from each bank are involved in the scheme. The left section of each bank row displays its details, including the total number of incoming and outgoing transactions and the total amounts sent and received, encoded in micro bar charts.

Each bank account appears as a node along the x-axis according to the order of its first transaction (regardless of direction). All nodes are equidistant and fill out all the available horizontal space.

All accounts are encoded in two-segment donut charts, whose slices represent the sum of all incoming (purple) and outgoing (teal) transactions. The length of each segment is proportional to the highest transaction volume of an account in the network, which corresponds to half a circle.

Victim (yellow) and mule (red) accounts have specific encodings and fraudulent transactions are always displayed in a red gradient.

Transactions flowing between accounts are encoded in pointed edges, whose thickness increases with the transaction value. These edges can represent one or multiple transactions if there are several movements in the same direction between two accounts.

To reduce visual clutter in more complex networks, we opted to curve the edges, providing a more organic display and reducing intersections and overlapping links.

Users can track transaction timing and distribution using the timeline below the graph, animate the transaction sequence with the “play flow” button, and filter the network or consult the legend using the top toolbar.

In smaller and most common networks, this tabular sequential structure graph seamlessly showcases all banks and accounts in the money laundering scheme. However, observation of real-life cases in high-fidelity mockups and the pre-production environment revealed that networks exceeding 15 accounts weakened the graph’s readability due to node overlap and edge congestion. Given the need for nodes to occupy unique columns and the limitations of a desktop screen, a 15-node limit prevents horizontal dragging, which might hide part of the graph.

\section{Grouping scenarios}
\label{sec:gs}

We established general criteria for node grouping, validated by the development team.

\begin{itemize}
\item Node grouping requires a minimum of 15 accounts. 
    
\item Grouping excludes victims, mules, and accounts with fraudulent transactions, as they are key to the investigation and have specific encodings. 

\item Grouping is restricted to accounts belonging to the same bank. 

\item Consecutive same-bank accounts with mutual transactions must remain separate to prevent circular edges.

\item Group amounts cannot exceed the largest amount in the graph (half a circle in a donut chart). This value determines the segment length for all nodes. To ensure the user experience in future interactions, the donut graph bars should be fixed to prevent resizing across the graph when ungrouping.
\end{itemize}

\begin{figure}
    \centering
    \includegraphics[width=1\linewidth]{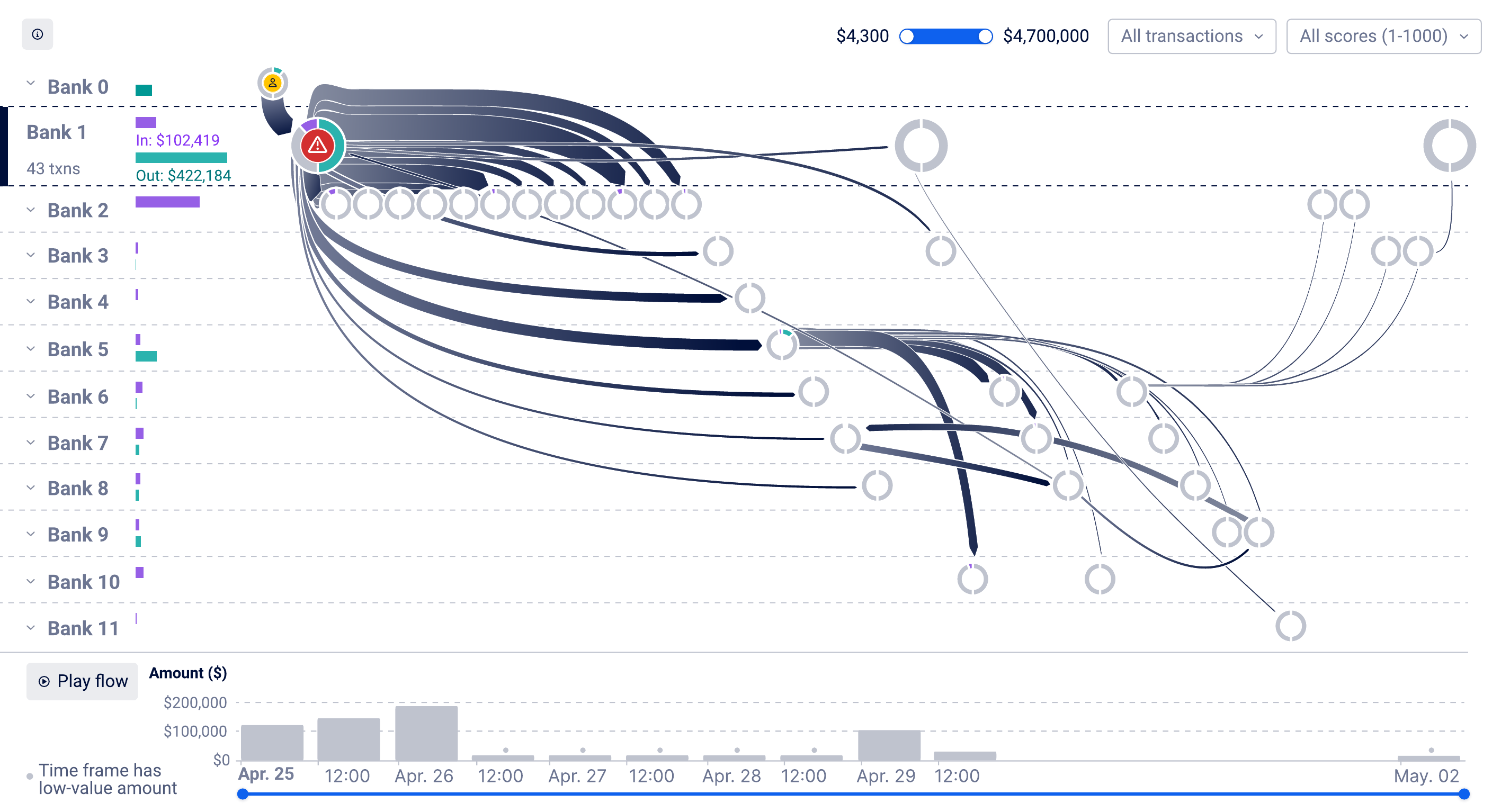}
    \caption{Example B ungrouped (synthetic data used for privacy).}
    \label{fig:examples_ungrouped}
\end{figure}

Three node grouping scenarios branched from these rules: one solely based on transaction amount, one based only on time constraints, and a combined solution that takes into account both the transaction amount and  order. 

We set thresholds at 5\%, 10\%, and 20\% of the largest network transaction for the amount-based and combined methods because transactions below 20\% represent around half of all transactions in real-life cases, reaching around 95\% of all transactions in the two examples in this study.
Based on domain knowledge, we established thresholds of 1, 12, and 24 hours for the time-based approach to fit money laundering schemes that vary from minutes to days. Initial explorations proved these intervals to be more adaptable than a proportional division of time.

We tested all scenarios and variants on two real money laundering cases. Example A (Figure ~\ref{fig:teaser}) had 45 accounts across six banks, and most transactions were routed to two banks over six hours. Example B (Figure ~\ref{fig:examples_ungrouped}) spanned over six days, 38 accounts, and 12 banks. All figures use synthetic data for anonymity but are structurally identical to the original examples A and B used throughout the research.

\begin{figure}
    \centering
    \includegraphics[width=0.75\linewidth]{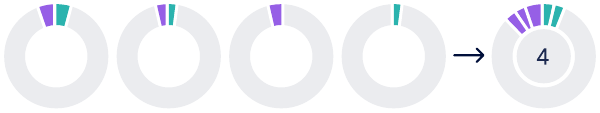}
    \caption{A meta-node encoding a group of four accounts.}
    \label{fig:meta_node}
\end{figure}

We also designed a new element: the meta-node (Figure~\ref{fig:meta_node}), a node of grouped accounts. As opposed to the single node, the meta node displays the cumulative incoming (purple) and outgoing (teal) transaction amounts for all accounts in that group and has a central circle showing the number of accounts.
In any grouping scenario, edges from grouped accounts are consolidated into a single edge if they share the same origin and destination.

\subsection{Amount and First Transaction Order Based Node Grouping (Combined)}

This approach preserves the essence of the tabular sequential network: drawing the nodes based on the order of the first transaction. Accounts are grouped if they are part of a sequence of transactions in which each incoming or outgoing amount is individually less than the set threshold (5\%, 10\%, or 20\%) of the largest transaction amount in the network. The grouping is interrupted by an amount greater than the threshold; a transaction to another bank in the network; or failure to meet a general criterion.Unlike the combined method, which stops grouping when transactions involve other banks, the amount-based approach prioritizes grouping sub-threshold accounts rather than preserving the network's original structure.

\begin{figure}
    \centering
    \includegraphics[width=1\linewidth]{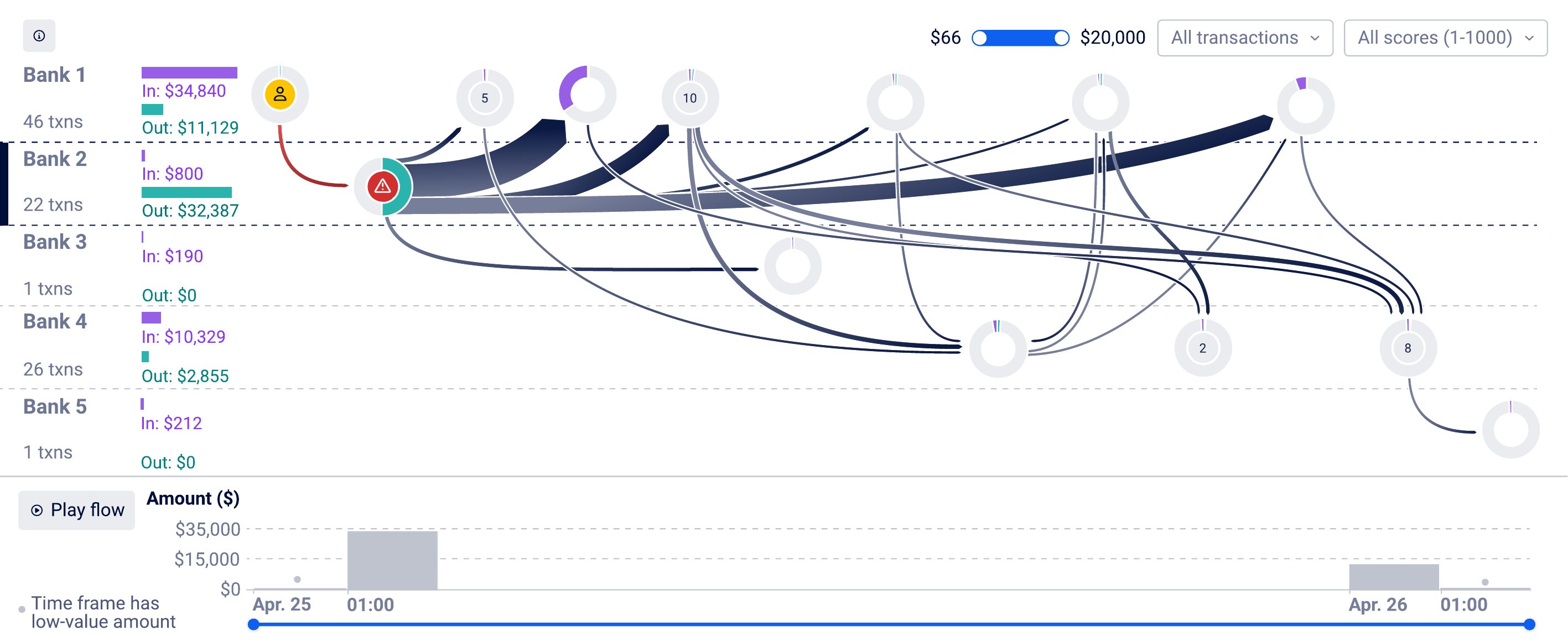}
    \caption{Combined node grouping in Example A with a 20\% amount threshold (synthetic data used for privacy).}
    \label{fig:combined_based_20}
\end{figure}

In Example A, we were able to reduce the number of nodes from 45 to 20 (in the 5\% and 10\% amount thresholds) and to 18 (in the 20\% amount threshold) (Figure~\ref{fig:combined_based_20}). This corresponds to a 55.6\% and 60\% reduction, respectively. In Example B, the change was less evident. The number of nodes fell from 38 to 31 at a 5\% threshold (an 18.4\% reduction) and to 29 at 10\% and 20\% thresholds (a 23.7\% reduction).

\subsection{Amount-Based Only Node Grouping}

In this scenario, the grouping only considers the transaction amount and discards the transaction order. That is, each account's transaction amount has to be individually below the threshold (5\%, 10\%, or 20\%) of the largest volume in the network. The meta-node is positioned on the first account that meets the criteria. The grouping is interrupted by an amount that exceeds the threshold or if it fails to meet a general criterion. 

This process results in a dramatic reduction in the number of nodes. In Example A, a network of 45 accounts now displays 9 nodes (with a 5\% amount threshold), which amounts to an 80\% decrease. In the 10\% and 20\% amount thresholds, the result is the same: a drop from 45 to 7 nodes (an 84.4\% drop) (Figure ~\ref{fig:teaser}, Grouped Network). 

Example B shows more modest node grouping: a decrease from 38 to 18 nodes (5\% amount threshold; an 18.4\% reduction) and from 38 to 29 nodes (10\% and 20\% thresholds; a 23.7\% reduction).

\subsection{Time-Based Node Grouping
}

The third grouping scenario is time-based and takes into account 1-hour, 12-hour, and 24-hour thresholds. Accounts are grouped if all incoming and outgoing transactions happen within the set intervals. The grouping is interrupted if the set time runs out or if a general criterion is broken. The meta-node takes the place of the node that first meets the criteria.

Using the 1-hour threshold, we were able to cut the number of nodes on the graph in Example A from 45 to 15, which corresponds to a 66.7\% decrease. This case occurred over a period of six hours. The 12-hour and 24-hour thresholds mirror the results and display of the 10\% and 20\% amount-based grouping, both showing an 84.4\% reduction from 45 to 7 nodes. 

\begin{figure}
    \centering
    \includegraphics[width=1\linewidth]{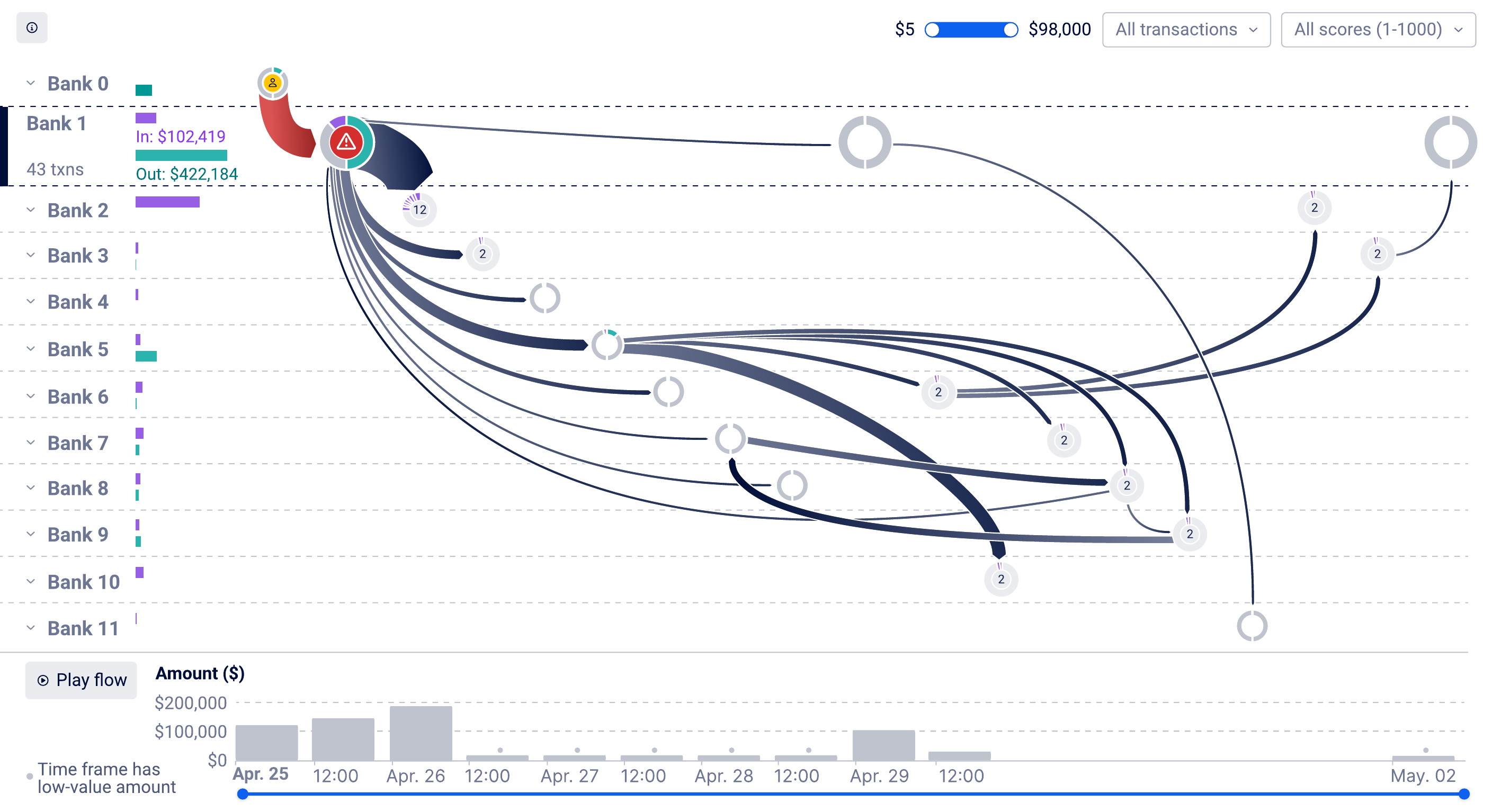}
    \caption{Time-based node grouping in Example B with 24-hours threshold (synthetic data used for privacy).}
    \label{fig:time_based_24hours}
\end{figure}

Conversely, as Example B spans nearly a week, the smallest threshold yielded few results. The 1-hour threshold reduced node count from 38 to 32 (a 15.8\% decline), marking the smallest reduction across all scenarios. The 12-hour and 24-hour (Figure ~\ref{fig:time_based_24hours}) thresholds are more effective. The 12-hour threshold reduces the node count from 38 to 24 (a 36.8\% decline), while the 24-hour threshold further reduces it to 19 (a 50\% reduction).

\section{User study}
\label{sec:us}

We conducted a qualitative user study to evaluate each node grouping scenario based on four key principles — readability, nuance (similarity to initial graph), visual clutter, and grouping structure. This involved 45-minute shadowed semi-structured interviews with three risk consultants specializing in money laundering, with a minimum of eight predefined questions across four sections—preference, design, expected interaction, and observations.
The questions addressed key fraud analysis tasks, such as pinning mule accounts, labeling transactions, and exploring account details, while evaluating how node grouping impacts the efficiency of these investigations.
Participants were given initial context and presented with four static versions of two examples with real-life data (A and B): the ungrouped network and the variant that yielded the most effective results in each scenario (20\% and 24-hour). For Example A, we chose the 1-hour time-based limit as the 12-hour and 24-hour thresholds produced results identical to the 20\% line in the amount-based scenario. We did a comparative analysis to prioritise and rank the users' preferences and insights and assess which could have a positive or negative impact on the  two key analytical tasks.

The amount-based scenario, which achieved the largest node reduction in both examples, was the unanimous choice for minimizing visual clutter and improving the readability of key accounts and fraudulent transactions. Users preferred it for its visual appeal, better focus, and significant reduction in on-screen elements.

Participants viewed the time-based approach positively, agreeing it best preserves the original networks. Unlike the amount-based scenario, which was deemed less interesting for analysis, the time-based method is considered more granular and similar to the network’s original flow. However, two analysts noted that this scenario would increase the analyst's workload due to the need to interpret more encodings and cross-check the timeline.

There was no consensus about the preferred grouping structure. Two consultants would rather keep some nuance preserved by the time-based and combined methods, while the remaining participant insisted that the solution with fewer elements on screen (amount-based) would be the most useful in real-life analysis. However, the analyst who chose the amount-based scenario stated that filtering, expanding, and using the surrounding widgets would be fundamental for a thorough analysis.

Participants validated the meta-node’s efficacy in visualizing groups. Despite positive feedback on its segmentation, the limited visibility of thin slices needs re-evaluating current minimum-width specifications in the deployed product.

The three consultants agree that node grouping effectively reduces visual noise, thereby streamlining the investigation process for complex cases.

The third set of questions focused on interactions: two interviewees expected meta-nodes to ungroup on click. All participants agreed that accessing detailed account information (via popovers, a side panel, or widgets) is essential for the analyst’s investigation.

\section{Discussion and conclusion}

In high-stakes fraud, analysts must prioritize speed and accuracy. Since networks quickly become complex, node grouping is essential to reduce visual clutter. This is critical for our proposed TSG visualizations of multi-bank money laundering cases, given their spatial constraints. 

We propose three node grouping methods for tabular sequential networks: combined, amount-based, and time-based. These methods target graphs with 15 or more accounts to improve the readability of key accounts (victims and mules) and fraudulent transactions, facilitating the analyst's investigation. All grouping methods are reproducible and adaptable by adjusting the threshold. The need for such alterations will be assessed with actual usage of this feature after its implementation and deployment.

The amount-based method showed the highest node reduction. But the combined and time-based approaches were were identified by expert users more efficient at capturing the network’s original structure while reducing clutter. 

The user study showed that node reduction alone is not a perfect solution as it removes nuance; two participants favored methods that preserved the original network structure. The trade-off for a slightly more crowded network is that while it may ease the manual examination of grouped nodes and the timeline, it requires added interpretation effort. Crucially, information for grouped accounts must be easily accessible for the analyst to proceed with the investigation, via summarized details, widgets, or an expandable network.

Since grouping methods produced uneven results in Examples A and B, we concluded that no single method is ideal for all money laundering schemes. However, the 24-hour time-based threshold shows the greatest potential for broader application, which we will propose as a system default. This option balances granularity for networks spanning multiple days with highly effective node reduction in briefer schemes, matching the amount-based results. 

For future work, we will harness the user study findings to prototype and implement node grouping and ungrouping interactions within the deployed tool.

A post-study revision of the grouping rules and scenarios showed that the time-based approach with the 12-hour and 24-hour thresholds, specifically when applied in Example A  and excluding fraudulent transactions, could further cut the number of nodes by 86.7\%.

\bibliographystyle{abbrv-doi}

\bibliography{NodeGrouping_Rev}
\end{document}